\documentclass[10pt,twocolumn,twoside,submit]{JCNtran}
\usepackage{color}
\usepackage{amsmath}
\usepackage{amssymb}
\usepackage{tabularx}
\usepackage{epsfig}
\usepackage{graphicx}
\usepackage[T1]{fontenc}
\usepackage{epstopdf}
\usepackage{cite}
\usepackage[utf8]{inputenc}

\def\BibTeX{{\rm B\kern-.05em{\sc i\kern-.025em b}\kern-.08em
    T\kern-.1667em\lower.7ex\hbox{E}\kern-.125emX}}

\begin{document}
\bibliographystyle{jcn}

\title{
IP Session continuity in heterogeneous mobile networks using Software Defined Networking}
\author{Petar D. Bojović, Živko Bojović, Dragana Bajić, Vojin Šenk
\thanks{This research was financially supported by the Ministry of Education, Science and Technological Development of the Republic of Serbia through Projects No. III 45003 and III 44003}
\thanks{P. D. Bojović aff. The School of Computing Union University Belgrade,  6/6 Knez Mihailova, Belgrade, Serbia}
\thanks{Ž. Bojović, D. Bajić, V. Šenk aff. Faculty of Technical Sciences University of Novi Sad,  6 Trg Dositeja Obradovića, Novi Sad, Serbia}
\thanks{P. D. Bojović is  the corresponding author (petar.bojovic@paxy.in.rs).}}

\markboth{JOURNAL OF COMMUNICATIONS AND NETWORKS}
{P. D. Bojović \lowercase{\textit{et al}}.: IP Session continuity in heterogeneous mobile networks using SDN} \maketitle

\begin{abstract}
Smart environment requires uninterrupted connection when moving from one network to another. This is best accomplished at the network level (L3). Full interoperability and integration of heterogeneous networks is necessary for communication session continuity. Software Defined Networking (SDN) with virtual IP addresses solves the problem. Implementing a homogeneous SDN is expensive, given the enormous investments in existing networks. To solve this second problem, we deploy the least set of SDN features to provide full L3 mobility. We use a common controller to manage the IP address translations.\end{abstract}


\vspace{10pt}
\section{\uppercase{Introduction}}
\label{sec:introd}
\PARstart{S}{mart} environments are the fastest growing modalities that deploy the communication networks. One of the major issues is to provide a strategy that enables full user mobility in IP based network \cite{AXM,KKP}. An example can be computer-to-vehicle integration that makes driving safe and comfortable \cite{YW}. 

Such an aim sets a series of challenges. One of them is to ensure the continuity of communication services \cite{CCQB} e.g. if a vehicle passes from a WiFi to a 3G network. Transfer from one to another coverage zone requires Layer 3 (L3) mobility, but, so far, no wider implementation of such a system exists. We present a solution for providing session continuity at network layer. It uses two concepts: software defined networking and IP address translation. This translation substitutes real temporal IP addresses with virtual permanent IP address. This transition results in seamless connection on both server and client sides. 

To ensure efficient implementation, the server should see the client as a static entity, even if it moves across the networks. In heterogeneous networks, NAT does not provide consistent and flexible network address masking. SDN, on the other hand, centralizes the masking procedure and keeps connection continuity. We propose to use virtual permanent address allocation to mobile user \cite{YLJZWV}. It solves L3 mobility problem and prevents interrupts and subsequent reconnections \cite{DBS}. 

Several solutions to this problem already exist, for L2, L3 and L5 mobility, systematized in \cite{ZWZ}. SDN technology at L2 layer was first applied in \cite{VKKHP,LMR}. In \cite{YLJZWV,WAAQM,QDGBV} SDN is explained in heterogeneous networks with increased programmability. SDN enables Internet mobility applying the ProxyMIP at L3 layer \cite{K} but such a solution requires installation of a software application provided by the network operator.

L3 mobility solutions that implement Proxy MobileIPv6 (PMIPv6) \cite{GLDCP}, require Access Point - Home router tunneling, to provide the mobile user the same IPv6 address while transferring from one to another MAG coverage regions. MAG to LMA tunneling eliminates the necessity for additional management services (software installation, end user tunneling interface). Unfortunately, introducing tunnels - i.e. virtual point-to-point MAG-LMA connections for each end user violates the optimal routing in PMIPv6 networks, while simultaneously adding supplementary tunnel overhead and reducing the payload. Thus the complete link efficiency is reduced \cite{GLDCP}.
 
A more efficient SDN solution for L3 mobility in \cite{WBZ,BOSBCJZ,TGVS} rewrites the L3 address within the packet. This solution proposes a complete migration to SDN network concept, which is too expensive. Namely, it requires a complete replacement of existing equipment with the SDN one, which is still under development and expensive. Such a complete replacement is complex and time consuming. 
\vspace{10pt}

The contribution of this research is to find a solution to L3 mobility that would:
\begin{itemize}
  \item Maintain the advantages of existing solutions considering the end user transparency (like PMIPv6)
  \item Combine the traditional L3 routing and L2 mobility with SDN packet flow management; such a functionality has a modest equipment requirements, so that the optimal routing and link efficiency are achieved with minimal investment
\end{itemize}

\vspace{10pt}
In this paper, we present a novel solution for Internet connection mobility problem, transparent both for the server and for the client in an IP network. The traditional IP network concepts that efficiently solve routing and traffic management problems are preserved, while SDN is implemented to address the problems beyond the traditional IP network capabilities, such as the client mobility. This paper considers the mobile network architecture that combines the best properties of the traditional mobility techniques L3 routing and L2 mobility of homogeneous access networks AS with the advantages of SDN packet flow management that would solve the L3 mobility problem translating the client address at L3 level.

\vspace{10pt}
\section{Architecture and Protocol Design}\label{sec:architecture}

\begin{figure}[h!]
\begin{center}
\includegraphics[width=8cm]{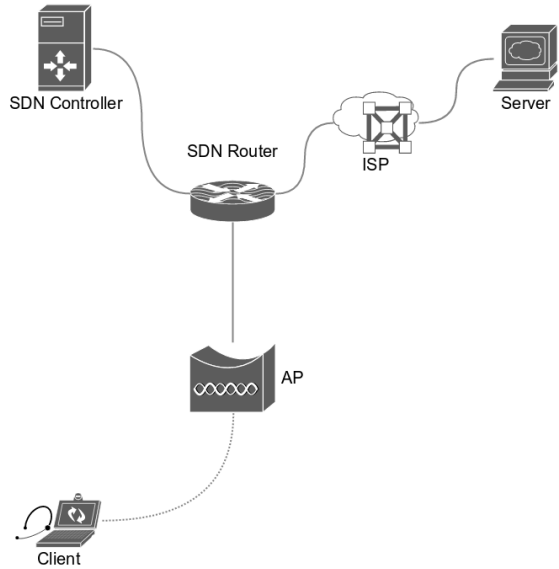} \caption{Basic SDN architecture.}\label{fig:sdnarch}
\end{center}
\end{figure}

The L3 mobility requires a change of client's IP address with each change of coverage zone \cite{GGZZ}. To prevent TCP connection interrupts, we need to emulate a permanent IP address. Traditional NAT cannot provide flexible and efficient mapping of private IP addresses for the needs of IP mobility, or for the same user in different networks. The NAT translates every single private address into a unique (different) public IP address, even for the same user. SDN, with a centralized controller, provides efficient and flexible management of masking tables. Figure \ref{fig:sdnarch} presents a possible SDN scenario for this purpose. The main idea of our research is to implement SDN for dynamic translations between real (rIP) and virtual (vpIP) IP address.
The advantage of this approach is that we eliminate involvement of administrator and reduce the cost and possibility of errors.

\begin{figure}[h!]
\begin{center}
\includegraphics[width=8cm]{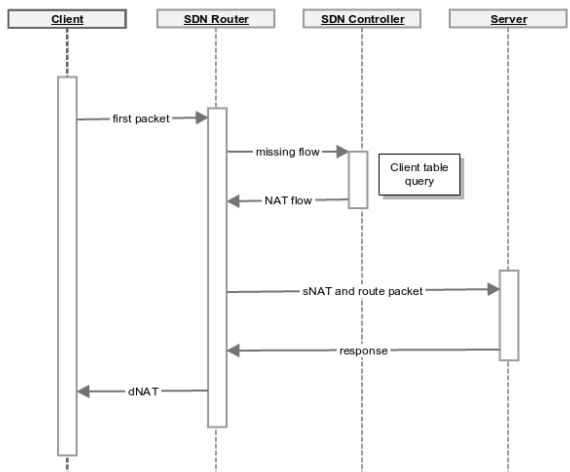} \caption{SDN packet flow diagram.}\label{fig:sdnflow}
\end{center}
\end{figure}

Figure \ref{fig:sdnflow} presents basic processes when packet is transmitted from a client to a server. A client sends packets with a real IP address A (rIP-A) to a server with an IP address B. When the first packet arrives to an SDN-aware device, it is immediately forwarded to an SDN controller. The controller updates the data tables that contain triplets of data for a mobile user (user identifier, such as MAC address, rIP and vpIP). Then it forms an open-flow instruction and forwards it to a router. The router translates the rIP into the vpIP and forwards the packet. At the server, the packets arrive with vpIP as a source address.

\begin{figure*}[h!]
\begin{center}
\includegraphics[width=\textwidth]{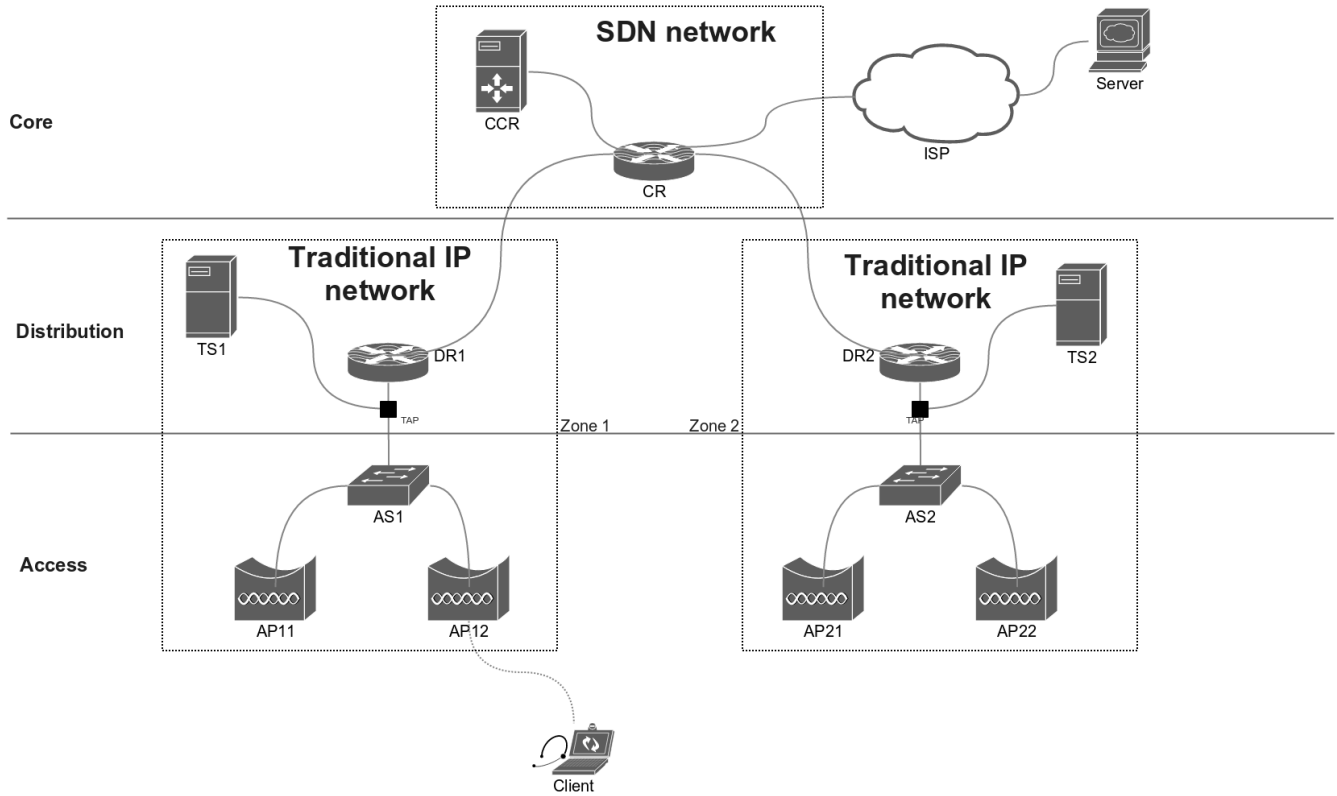} \caption{Proposed mobile network architecture.}\label{fig:arch}
\end{center}
\end{figure*}

The proposed architecture is presented in Figure \ref{fig:arch}. It can cover wide areas and, basically, it comprises a three-level hierarchy (core, distribution and access) that should increase the network scalability and reliability. rIP translates to vpIP at the core level, using the controllers for core router (CCR). But the problem with CCR is that it cannot identify users according to the MAC address, since this information is located at the lower layers. Introducing a Tap Server (TS) to collect packets between access and distribution layer resolves the problem. Each coverage zone has its own TS that collects the information on MAC and rIP client addresses and forwards it to CCR. TS periodically scans the client and informs the CCR that the client is present. TS has no role within the routing process. When the client changes the coverage zone, new TS informs the CCR about the change and forwards the client's universal identifier and his new rIP. The CCR updates the table, but, only rIP has changed out of the data triplet that corresponds to the client. Since mapping of the universal identifier to vpIP remains the same, the connection is not interrupted during the change. Layer 2 procedures at access level remain the same as L2 roaming procedures.

The architecture proposed in Figure \ref{fig:arch}. introduces minimal changes into existing IP networks. They are introduced at the core layer by implementing SDN functionality. The rest of the network is implemented using traditional approach.

\vspace{10pt}
\section{Implementation and Evaluations}\label{sec:eval}
\subsection{Testbed Environment}

\begin{figure}[h!]
\begin{center}
\includegraphics[width=8cm]{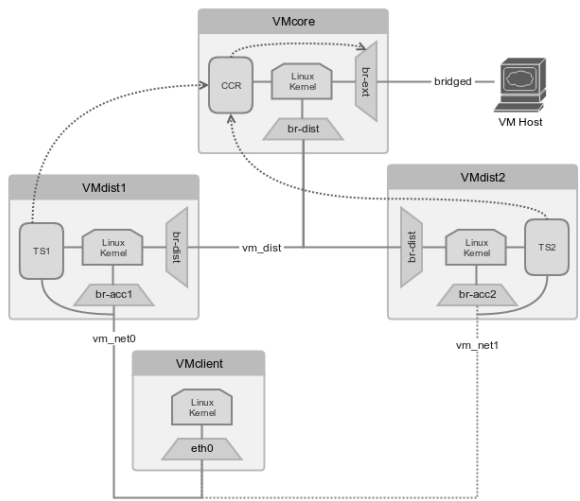} \caption{Test environment.}\label{fig:test}
\end{center}
\end{figure}

Figure \ref{fig:test} presents a test model that comprises four virtual machines, operating at different hierarchical levels of the proposed model:

\begin{itemize}
\item Vmcore, operating at the core level,
\item Vmdist-1 and Vmdist-2, operating at the distribution level,
\item Vmclient, simulating the wireless client. 
\end{itemize}

\vspace{10pt}
The virtual client machine for L3 roaming connects to the virtual machine at the distribution level using Ethernet. SDN controller integration and the multilayer Open vSwitch (OVS) technology for network management via OpenFlow protocol both enable higher programmability.

Each hierarchical level has two virtual OVS bridges. They communicate with lower and higher levels of the proposed network model. The bridges have static IP addresses, while their mutual interconnection uses Linux kernel. Besides the traditional switching functions, the border OVS bridges are responsible for external network communications (br-ext bridge connected to the link towards the provider network), as well as for access level clients communication (br-acc, connected to the client machine), thus enabling the intelligent network flow management. Connection of bridges to controllers enables dynamic flow control of the routes toward both the external network and towards the client. OVS bridge communication to the controller uses the OpenFlow protocol v1.3, supporting both IPv4 and IPv6.

The Floodlight SDN controller implements the network control logic. It includes MobilityService module for core level controllers. 

\begin{figure*}
\begin{center}
\includegraphics[width=\textwidth]{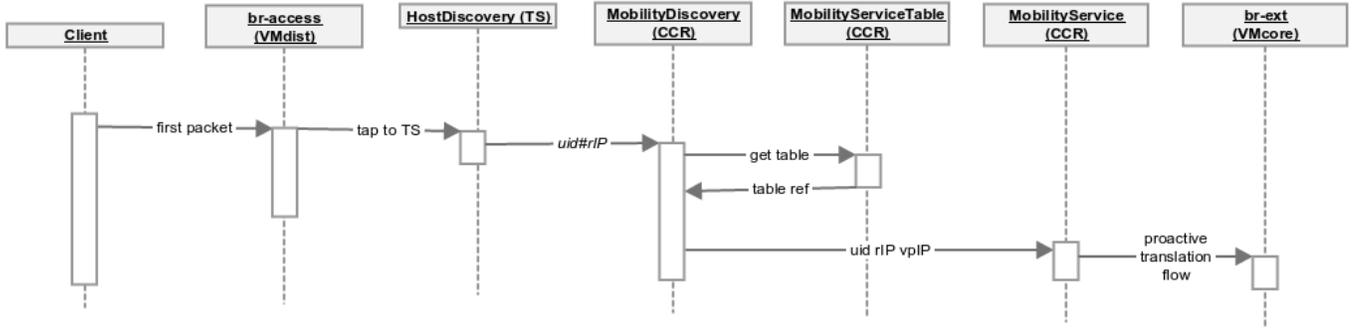} \caption{UML flow diagram of packet processing.}\label{fig:flowdiagram}
\end{center}
\end{figure*}

UML flow diagram in Fig. \ref{fig:flowdiagram} shows procedures performed by the Mobility service module. They define the proactive translation flows for br-ext bridge and enable L3 mobility.

\subsection{Methodology}
The core level controller initiates the MobilityDiscovery (MD) thread. A part of MobilityService module, it handles new connections arriving from the TS at the distribution level. When distribution level TS allocates a new client, it establishes a socket communication to MD. MD, in turn, provides an information in a form uid\#rIP (uid=userId, e.g. MAC). Here, \# stands for concatenation of two sequences. MD processes it, checking whether the MobilityServiceTable (MST) already contains the uid record. In the case the record is absent, MD decides that a new client has arrived and allocates a new random vpIP address. It implies creating a new table record that consists of uid, rIP, and vpIP fields. It also creates new proactive flows for sNAT (source NAT, rIP into vpIP translation) and dNAT (destination NAT, vpIP into rIP translation). For this, it uses parameters from the table (fig. 5). The br-ext bridge receives it with a time limit, after which it cleans up the flow record at br-ext. The new flow record must have a higher priority than the one for L2 packet processing.

At the core level, MAC address is not available, having a functional distance from the user L2 link. With a new HostDiscovery (HD) module implemented inside TS at the distribution level, this is not a problem. It collects information about user identifications uid and their respective rIP. Hardware tap between access and distribution layer takes care of getting TS sample of every user packet. Hardware tap is a hub that transfers each packet of a particular link to TS for further analysis. Overhead analysis of a packet increases the reliability of the proposed method for a new client search.  An alternative to a hardware tap is a software tap (Port Mirror). To avoid an additional load that a software tap may cause to existing routers, we opted for a hardware tap. To decrease the TS load, only DHCP or Router Solicitation packets could be analyzed, thus reducing the procedure.  

HD checks the source IP address of each packet, comparing it to the range expected. This protects against spoofing and packet rejection before the client obtains a correct rIP address. E.g., a client is within a DHCP process and his source address is 0.0.0.0. When HD receives a packet with a correct riP address, it forms a socket towards CCR, proceeding a message in a form uid\#rIP (in our example, MAC\#rIP). For each new client arriving to CCR, HD inserts an ordered pair (rIP, SystemTimeInMiliseconds) into a Time Buffer. Every x minutes TS sends a message to CCR to inform about clients present and updates the time in the Time Buffer (update time is fully configurable). Update time period of 5 minutes was found to be optimal during simulation testing. 

Without additional requirements or packets, virtual machine (VM) provides transparency to the client. It only requires a correct functionality of DHCP client during the network change. We performed the test model (Fig. 4) simulation study using the Oracle VirtualBox virtual machines. VirtualBox establishes a virtual network configuration creating the fundamentals of the required scenario. We created a virtual machine for each Access level within different zones and between the distribution and core levels. A quick change of networks simulated the mobility (L3 roaming). It used the VirtualBox command enabling a virtual network change of a particular VM. A script shown in Tab. \ref{table:cmds}. realizes transfer from zone 1 to zone 2. The script contains the VirtualBox commands to
\begin{itemize}
	\item disable the first network adapter of VM Client;
	\item connect the first network adapter to virtual network vm\_net1; 
	\item enable the first network adapter.
\end{itemize}
\vspace{10pt}

The same procedure returns to the zone 1, except for the virtual network name that would be vm\_net0.

\begin{table}[h!]
\begin{center}
 \begin{tabular}{||c||} 
 \hline
 Commands \\ [0.5ex] 
 \hline
 VboxManagecontrolvm "Client" setlinkstate1 off \\ 
 \hline
 VBoxManagecontrolvm "Client" nic1 intnet vm\_net1 \\
 \hline
 VBoxManagecontrolvm "Client" setlinkstate1 on \\
 \hline
\end{tabular}
\caption{The script that simulates the zone change}
\label{table:cmds}
\end{center}
\end{table}

\vspace{10pt}
\section{Evaluation of Results}\label{results}	
After client connects to an access node in zone 1, DR1 allocates it a dynamic IP address. After sending the first packet, DR1 processes it according to the routing table. Also, it is received by TS1's HostDiscovery module. HD forwards the information about both the allocated rIP of zone 1 and the client's MAC address (uid\#rIP) to the MobilityDiscovery at CCR. MD initiates the mapping table updates, and it sends the records to MST. MST checks if the client exists. If not, MST allocates vpIP and stores it, together with client's data, into the mapping table. It also sends, using Mobility Service, the proactive flows to CR for sNAT and dNAT translations. If CR receives a packet with rIP before updating the proactive policy, CR retrieves the translation instructions. CR translates the packets with rIP into vpIP and vice versa, according to flow tables and without controller interaction. Server at the external network observes the communication via the vpIP address only, and CR translates each rIP address into vpIP. HD periodically informs CCR that client is still using the network, while CCR updates the flows with a time limit.

Suppose a client transfers from zone 1 to zone 2. In our scenario, a network changing script initiates a processes in which client requests and gets a new zone 2 dynamical IP address. DR2 then processes the packet according to its routing table. TS2's HD accepts the information about the new user and sends it to MD at CCR. MST modifies rIP address to the IP from the new zone, keeping the other table parameters (uid and vpIP) unchanged. MST also, through MS, sends to CR the flows to translate new rIP into the allocated user's vpIP. The old translation remains active until time limit expires, but since there are no packets, no problems occur. The flow of the new translation would continue mapping the new rIP into existing vpIP. The final outcome is that the server at the remote network perceives all the communication as coming from one identical vpIP.

\begin{figure}[h!]
\begin{center}
\includegraphics[width=8cm]{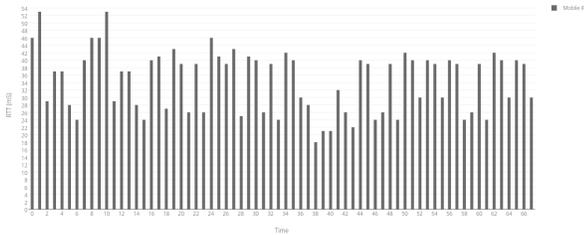} \caption{TCP communication from server side.}\label{fig:tcpserver}
\end{center}
\end{figure}
\begin{figure}[h!]
\begin{center}
\includegraphics[width=8cm]{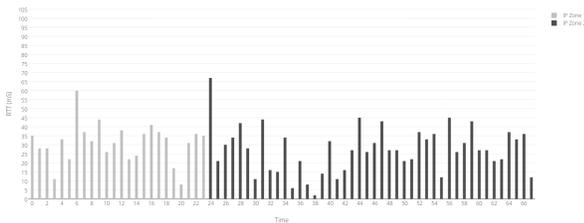} \caption{TCP communication from client side.}\label{fig:tcpclient}
\end{center}
\end{figure}

Figures \ref{fig:tcpserver} and \ref{fig:tcpclient} show recoded RTT (round trip time) values, for the same TCP communication, from server and client points of view.   

\begin{figure}[h!]
\begin{center}
\includegraphics[width=8cm]{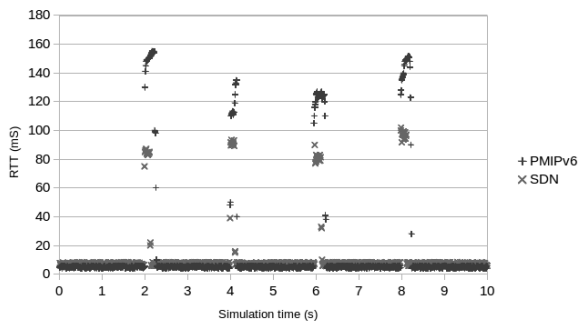} \caption{Switch-over packet delay}\label{fig:packetdelays}
\end{center}
\end{figure}

To confirm our solution we have compared our results with Proxy Mobile IPv6 (PMIPv6) \cite{GLDCP} solution. For this reason, we have used Mininet implementation of PIMPv6 and developed our solution in Mininet environment. Figure \ref{fig:packetdelays}. shows results of RTT for these two solutions. Data show improved performance of IP mobility (Shorter time for switching from a private IP address of one network to a private IP address of another network) with SDN solution as compared to PIMPv6 approach.

\begin{figure}[h!]
\begin{center}
\includegraphics[width=8cm]{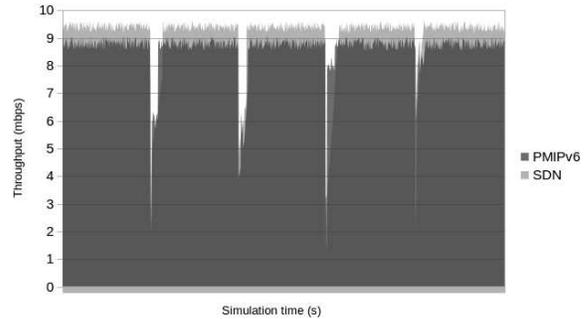} \caption{Switch-over packet throughput}\label{fig:throughput}
\end{center}
\end{figure}

Maximal throughput is also tested on 10mbps links (Figure \ref{fig:throughput}). Tunnel encapsulation causes visible decrease of throughput in  PMIPv6. Decrease in throughput of the proposed method is due to processing time before correct NAT flow appears on CR.

Consistency in IP addressing is a crucial point in L3 mobility. It enables undisturbed TCP-protocol services and prevents interrupts. The established method, verified by simulation, is perfect for this goal (Fig \ref{fig:tcpserver} and \ref{fig:tcpclient}). There is neither packet loss, nor connection failure during the roaming process.
The results are obtained using laboratory equipment and with virtual machines. The follow-up research would include the implementation of the proposed method in a real environment.

\vspace{10pt}
\section{CONCLUSION}\label{sec:conclusions}
Traditional IP networking cannot solve the L3 mobility problem. SDN can solve it efficiently. We have proposed to enable L3 mobility translating the client address at L3 level. It enables continuous data flow, which is of uttermost importance for Smart Things concept. Seamless IP mobility in heterogeneous networks is crucial here. 

Compared to PMIPv6, our proposal does not use tunneling, but original packets with least changes at the central router. As clients packets are not changed inside local domain, optimal local routing and maximal link efficiency is preserved.

Implementing a homogeneous SDN is expensive, given the enormous investments in existing networks. To solve this second problem, we deploy the least set of SDN features to provide full L3 mobility (one SDN controller and one SDN switch per network). We use a common controller to manage the IP address translations. Future research should yield a solution for this problem without further hardware investment.

\vspace{10pt}
\bibliographystyle{IEEE}

\epsfysize=3.2cm
\begin{biography}{pbojovic}{Petar D. Bojović} has graduated with Master degree on Faculty of Computer science on Union University Belgrade in 2008. In June 2008, he joins Faculty of Computer science as Lecturer in department of computer networks. Presently he works as Associate Professor at Faculty of Computer science on teachings and research. His interest includes Computer networks and Security of computer network. Currently, he works on doctoral thesis on area of computer network security and SDN.
\end{biography}
\epsfysize=3.2cm
\begin{biography}{zbojovic}{Dr. Živko Bojović}
  is assistant professor on the courses covering computer and telecommunication networks, IP technology, software of telecommunication systems and storage infrastructure and communications in Big data. He received the Ph.D. degree in Electrical Engineering from the University of Novi Sad in 2011 and joined the Communication Engineering and Signal Processing Chair in 2015. Before his academic career he worked as director of Department for Construction and Maintenance of Business and Technical Buildings in Logistics Division of ``Telekom Srbija``. He was participant and project leader for a number development projects financed by Telekom Srbija.
\end{biography}
\epsfysize=3.2cm
\begin{biography}{dbajic}{Prof. Dr. Dragana Bajić} is professor of the Faculty of Technical Sciences, University of Novi Sad. In 1995 she received the annual award for best Ph.D. thesis from the Commercial Chamber of Belgrade. She is the author of a number of scientific papers published in leading journals and at renowned symposia. Her research interests include frame synchronization, stochastic processes in communications, biomedical data processing, and error control coding.
\end{biography}
\epsfysize=3.2cm
\begin{biography}{vsenk}{Prof. Dr. Vojin Šenk} is the head of Communication Engineering and Signal Processing chair at the Faculty of Engineering (aka Technical Sciences) of the University of Novi Sad.  He gained his bachelor, masters and PhD degrees, all in electrical engineering, from the universities of Novi Sad (BsC) and Belgrade.  His research interests are in the area of Information and Coding Theory.  He has published more than 170 scholarly papers in journals and conferences. 
\end{biography}

\end{document}